\documentclass[11pt,a4paper]{article}

\usepackage[utf8]{inputenc}
\usepackage[T1]{fontenc}
\usepackage{geometry}
\usepackage{tikz}
\usetikzlibrary{fadings}
\usepackage{xcolor}
\usepackage{graphicx}
\usepackage{lipsum} 
\usepackage{fancyhdr}
\usepackage[hidelinks]{hyperref}
\usepackage{amsmath}
\usepackage{amssymb}
\usepackage{enumitem}
\usepackage{titlesec}
\usepackage{longtable}
\usepackage{booktabs}
\usepackage{parskip}
\usepackage{pdfpages}
\usepackage[numbers,sort&compress]{natbib}
\usepackage{url}

\usepackage[T1]{fontenc}
\usepackage{roboto}

\title{Identity Management for Agentic AI: \\ {\Large The new frontier of authorization, authentication, \\ \vspace{-0.5em}
and security for an AI agent world}}
\author{Tobin South \and Subramanya Nagabhushanaradhya \and Ayesha Dissanayaka \and Sarah Cecchetti \and George Fletcher \and Victor Lu \and Aldo Pietropaolo \and Dean H. Saxe \and Jeff Lombardo \and Abhishek Shivalingaiah \and Stan Bounev \and Alex Keisner \and Andor Kesselman \and Zack Proser \and Ginny Fahs \and Andrew Bunyea \and Ben Moskowitz \and Atul Tulshibagwale \and Dazza Greenwood \and Jiaxin Pei \and Alex Pentland}
\date{}

\hypersetup{
    pdftitle={Identity Management for Agentic AI},
    pdfauthor={Tobin South},
    pdfsubject={The new frontier of authorization, authentication, and security for an AI agent world},
    colorlinks=true,
    linkcolor=black,
    citecolor=black,
    filecolor=blue,
    urlcolor=blue,
}

\definecolor{gradientstart}{RGB}{225,224,166}
\definecolor{gradientend}{RGB}{121, 168, 248}
\definecolor{odiforage}{RGB}{235, 124, 4}
\definecolor{workospurple}{RGB}{99, 99, 241}
\definecolor{textlight}{RGB}{245,245,250}

\begin{document}

\includepdf[pages=1]{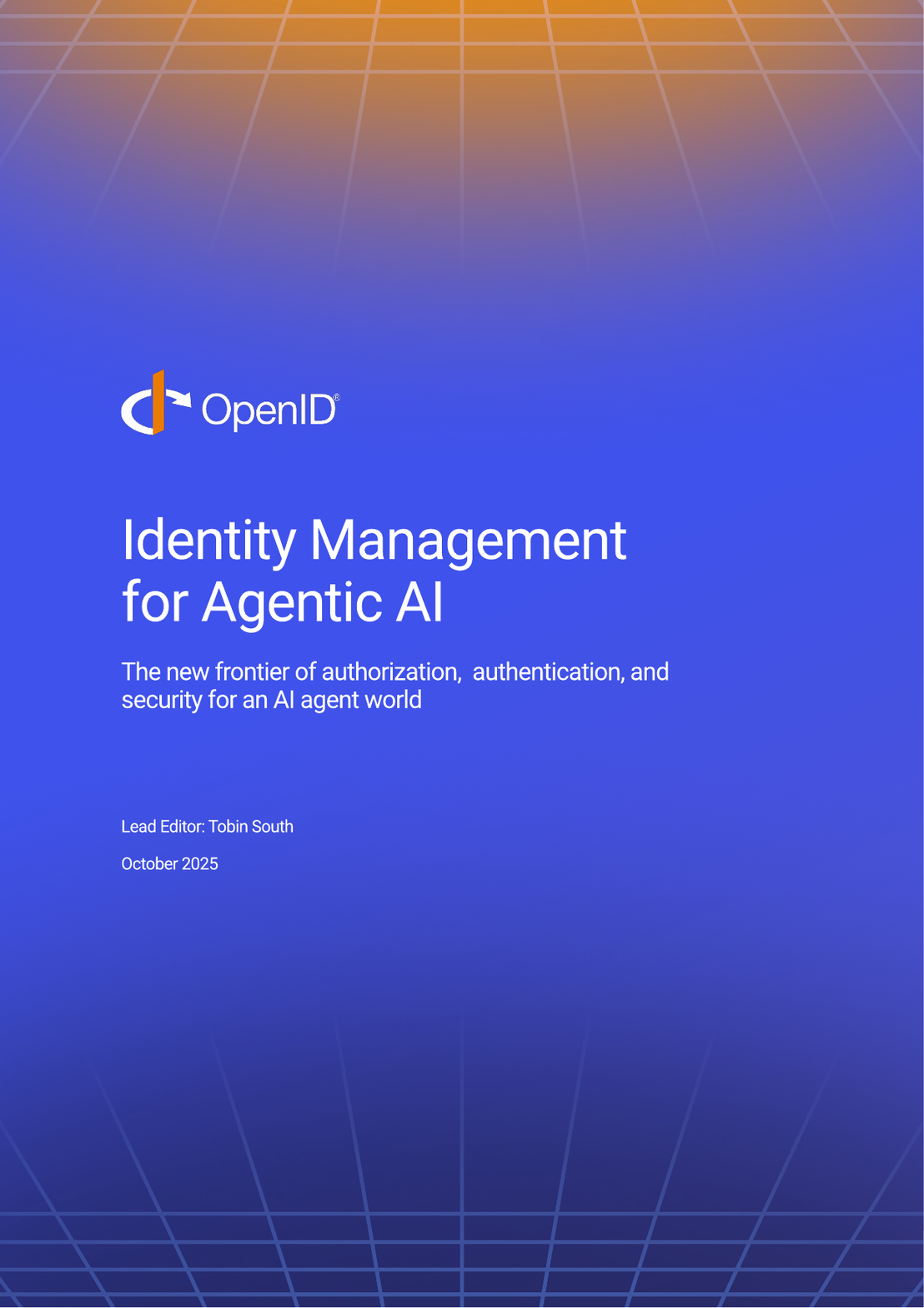}

\pagenumbering{arabic}
\setcounter{page}{1}

\section*{Executive Summary}

The rapid rise of AI agents presents urgent challenges in authentication, authorization, and identity management. Current agent-centric protocols (like MCP) highlight the demand for clarified best practices in authentication and authorization. Looking ahead, ambitions for highly autonomous agents raise complex long-term questions regarding scalable access control, agent-centric identities, AI workload differentiation, and delegated authority. This whitepaper is for stakeholders at the intersection of AI agents and access management. It outlines the resources already available for securing today's agents and presents a strategic agenda to address the foundational authentication, authorization, and identity problems pivotal for tomorrow's widespread autonomous systems.

\subsection*{Today's frameworks handle simple AI agent scenarios:}

\begin{itemize}
\item \textbf{AI agents differ fundamentally from traditional software;} they take autonomous actions on external services, exhibiting non-deterministic, flexible behavior that adapts in real-time, rather than simply executing predetermined instructions.  
\item \textbf{Existing OAuth 2.1 frameworks, when used with AI agents, work well within single trust domains} with synchronous agent operations (e.g., enterprise agents accessing internal tools, consumers accessing their services through AI tools), but may fall short in scenarios that are cross-domain, highly autonomous, or asynchronous, as well as those which require the agent to use or enforce delegated permissions on behalf of multiple human users at the same time.  
\item \textbf{The Model Context Protocol (MCP) is leading in adoption} as the key framework for connecting language models to external data sources and tools when building agents. Other approaches exist and should be supported, including function calling (tool use) and agent-to-agent (e.g., A2A) communication protocols.   
\item \textbf{Enterprise SSO and SCIM provisioning can help enable the use of enterprise agents} and facilitate centralized agent lifecycle management, as well as governance of permissions and access for various AI agent use cases.  
\item \textbf{Enterprise security profiles provide the baseline for safe AI adoption.} To mitigate risks, this paper recommends that AI agents conform to rigorous, interoperable profiles of existing identity standards. Working groups, such as the Interoperability Profiling for Secure Identity in the Enterprise (IPSIE), can provide guidance on this, giving organizations the confidence to adopt AI by ensuring robust controls are in place.  
\item \textbf{User-centric consent models are the foundation for consumer agents.} For agents connecting to third-party consumer services (e.g., email, social media, financial data), the established OAuth 2.1 user consent flow is the primary mechanism for granting delegated authority, making transparency and clear scope definition critical for user trust.
\end{itemize}

\restoregeometry
\clearpage

\subsection*{Critical future challenges exist:}

\begin{itemize}
\item \textbf{Agent identity fragmentation should be avoided.} Vendors could develop proprietary agentic identity systems, which would reduce developer velocity by forcing repeated one-off integrations. It would also compromise security by creating multiple security models, each with different risks and vulnerabilities.  
\item \textbf{User impersonation by agents should be replaced by delegated authority.} Currently, agents often act indistinguishably from users, creating accountability gaps and security risks. True delegation requires explicit ``on-behalf-of'' flows where agents prove their delegated scope while remaining identifiable as distinct from the user they represent.  
\item \textbf{Scalability problems exist in human oversight \& user consent.} Users will face thousands of authorization requests as agents proliferate, creating security risks from reflexive approval. Preemptive authorization and scoping of flexible agents are at odds with least privilege.  
\item \textbf{Recursive delegation creates risks.} Agents spawning sub-agents or communicating tasks to other agents create complex authorization chains without clear scope attenuation mechanisms.  
\item \textbf{Agents acting on behalf of and reporting to teams of humans lack support.} While OAuth and OpenID Connect were designed for individual user authorization, agents can be employed in shared codebases or chat channels in groups. In these multi-user environments, various permission levels may exist for different users, but all of them share a common objective within a single context. No popular protocol exists to support shared agents.  
\item \textbf{Trustworthy autonomy lacks automated verification.} Scaling beyond human-in-the-loop safety models requires new, programmatic methods to ensure an agent's actions continuously align with its operational goals and constraints.  
\item \textbf{Browser and computer-use agents break the current authorization paradigm.} Agents controlling visual interfaces directly (or via MCP into browser orchestrators) bypass all traditional API-based authorization controls. Protecting the open web from lockdown will require robust authentication of web bots or web agents.   
\item \textbf{Multi-facet behavior of agents complicates identity.} Technological advancements allow an agent to act on its own, requiring agents to have their own credentials, permissions, and audit trails. Furthermore, an agent's nature can be hybrid, enabling it to alternate between independent execution and acting on behalf of a user.
\end{itemize}

\clearpage



\setcounter{tocdepth}{2}
\tableofcontents
\clearpage

\maketitle
\section*{Contributors}

This report was prepared for the OpenID Foundation by Tobin South starting in April 2025, in collaboration with the \href{https://openid.net/cg/artificial-intelligence-identity-management-community-group/}{Artificial Intelligence Identity Management Community Group}, Stanford's \href{https://loyalagents.org/}{Loyal Agents Initiative}, and many independent reviewers and contributors, with thanks to WorkOS for supporting its completion with the author's time and practical contributions towards testing the ideas herein.
\\

Many collaborators, co-authors, advisors, and reviewers contributed to this work. Special thanks for direct contributions goes to Subramanya Nagabhushanaradhya, Sarah Cecchetti, Ayesha Dissanayaka, George Fletcher, Aaron Parecki, Pamela Dingle, Tom Jones, Aldo Pietropaolo, Łukasz Jaromin, Tal Skverer, Wils Dawson, Andor Kessleman, Stan Bounev, Kunal Sinha, Bhavna Bhatnagar, Nick Steele, Abhishek Maligehalli Shivalingaiah, Victor Lu, Chris Keogh, Govindaraj Palanisamy (Govi), Alex Keisner, Mahendra Kutare,  Pavindu Lakshan, Michael Hadley, Zack Proser, Cameron Matheson, Dan Dorman, Zack Prosner, Alex Pentland, Jiaxin Pei, Dazza Greenwood, Ginny Fahs, Andrew Bunyea, Ben Moskowitz, Atul Tulshibagwale, Jeff Lombardo, Elizabeth Garber, and Gail Hodges. 

\newpage
\section{Agents are Different this Time?}

\textit{The first step in understanding the unique identity, authentication, authorization, and audit needs of AI agents is to define exactly what they are and why they differ.} 

\begin{figure}[h!]
\centering
\includegraphics[width=\textwidth]{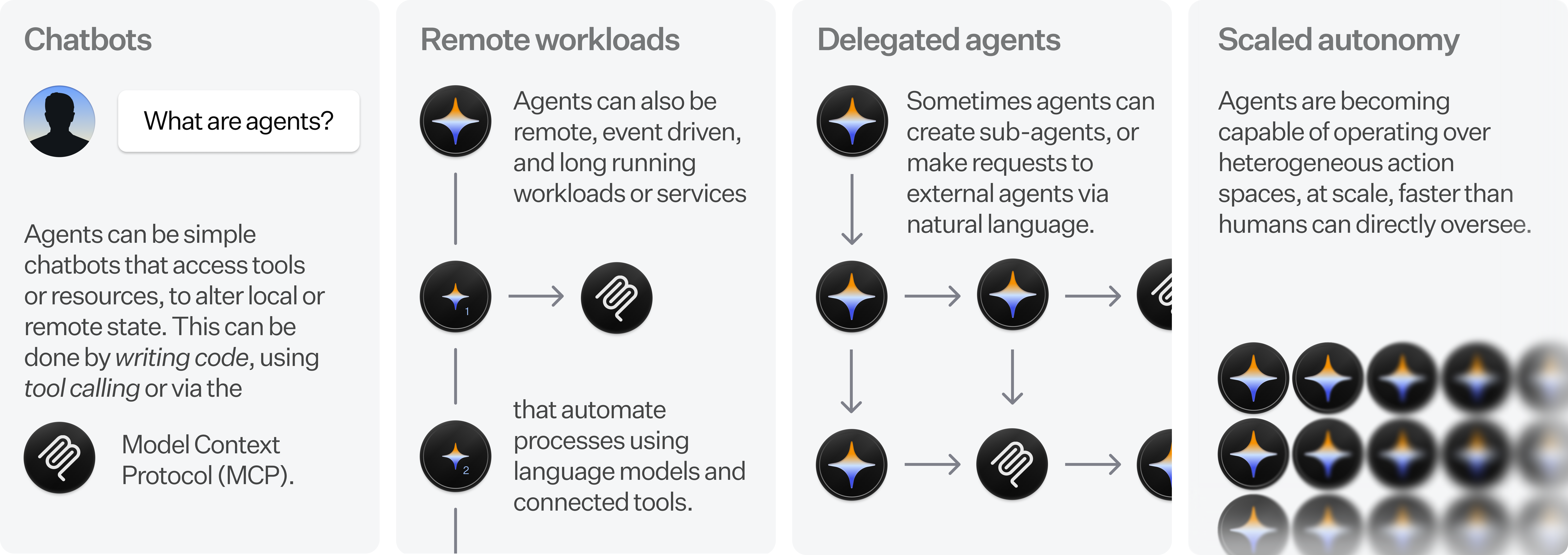}
\caption{An illustrative example of the different types of AI agents and how they use tools, such as the Model Context Protocol (MCP), across increasing levels of autonomy.}
\label{fig:agent-types}
\end{figure}

\subsection{Defining AI Agents}

AI systems are taking the world by storm and come in a range of formats, from chat interfaces using language models to internal workflow automations using traditional machine learning techniques. What delineates an AI \textit{agent} in the context of this paper is the \textbf{ability for the AI-based system to take `action' based on `decisions' made at model inference time to achieve specific goals}. This is beyond simple chatbots producing text outputs.  Unlike traditional software, which follows predefined rules and instructions, AI agents can learn from context, adapt to new situations, and make decisions autonomously. Traditional software typically requires manual updates to handle new scenarios, while AI agents use context and reasoning to improve their performance and adapt. They can interact via APIs as well as specialized agent communication protocols, such as the Model Context Protocol (MCP), or browser interfaces. Some of these behaviors can resemble standard remote workloads or applications, but are distinct due to their highly flexible, non-deterministic, and contextual nature.

Generally, ``agents'' are defined as identified and authorized software that uses language models to interact with external resources. Throughout this piece, examples of agents to consider include:

\begin{itemize}
\item Language model-backed chat interfaces that call specific AI-centric tools via MCP.  
\item Remote workflow automations using language models where system inputs result in tool use calls and database interactions (i.e., a persistent collection of software and language models together in a workload with a repeated but non-deterministic behavior).  
\item Semi-autonomous chain-of-thought style agents, which undertake a series of sequential steps and alternate between ``thinking'' in text and making requests to external tools or databases.
\end{itemize}

While many AI systems exist, this paper primarily focuses on large foundation model-based agentic systems (e.g., ChatGPT, Claude, Gemini, LLaMa, and systems built atop these models that allow them access to external tools), given the current progress and investment in this technology. More inclusive definitions of agents exist, such as AI systems focused on web search or computer-use agents that interact directly with a computer. These are important use cases worthy of discussion, but out of the scope of this paper to maintain a clear focus.

A final, critical distinction lies in the interaction paradigm itself. Traditional software clients--web, desktop, or mobile--operate on structured, unambiguous user inputs such as button clicks, form submissions, or menu selections. These actions represent a clear, auditable grant of intent. AI agents, by contrast, are designed to interpret unstructured and multimodal inputs. A user may provide instructions not only through text but also via documents, images, audio recordings, or video files. For example, a user might upload a scanned invoice image with a voice command, or they may forward a complex email thread with an instruction. This places a significant interpretive burden on the agent to extract semantic meaning, identify the scope of delegated authority, and formulate an execution plan from data that lacks the explicit, machine-readable consent signals of traditional UIs. This ambiguity at the point of instruction is a primary driver for new authentication and authorization models.

\subsection{Agents Need Specific Authentication and Authorization}

All online workloads require authentication and authorization--and agent-based workloads are no exception. However, agents are becoming increasingly autonomous, capable of engaging in multi-step processes while interacting with multiple external tools in sequence. This leads to unique concerns around user consent, specific regulatory implications (e.g., human-in-the-loop requirements), agent governance, and the granularity of access controls in large-scale and dynamic contexts. Agents represent a very specific set of workloads from the view of services, and there are important implications for how they should be handled, which will be outlined throughout this paper. 

Furthermore, as the number of AI agents and specific external resources grows and as the complexity of agent operations increases, manual or granular authorization of AI agent interactions will become untenable. Designing scalable and robust systems of trust, governance, and security will be crucial for achieving discovery, registration, and authorization at this scale.

The highly flexible, non-deterministic, and externally resource-connected nature of agents presents specific challenges to agent identity, governance, authentication, authorization, least privilege, audit, and other related aspects. This flexibility, driven by the need to interpret complex, unstructured inputs and interact with a dynamic set of external resources, presents specific challenges to agent identity, governance, authentication, authorization, least privilege, and audit. These challenges and the questions they raise are the focus of this white paper.

\subsection{Intended Audience}

This white paper addresses the distinct but interconnected security and identity challenges faced by the three primary groups shaping the AI agent ecosystem:

\begin{itemize}
\item \textbf{For AI implementers \& architects:} It provides technical guidance on using foundational standards like OAuth 2.1 and emerging protocols like MCP to build secure and interoperable agentic systems from the ground up.  
\item \textbf{For enterprises \& security leaders:} It outlines strategies for governance, risk, and integration, detailing how to apply familiar models like SSO and SCIM to manage the agent lifecycle and ensure compliance.  
\item \textbf{For consumer platforms \& product leaders:} It focuses on the foundations of user trust, tackling the core challenges of scalable consent, true delegated authority, and designing agent interactions that are transparent and safe for end-users.
\end{itemize}

\textbf{Section 2 of this report summarizes the current state of agent identity and authorization, pointing towards relevant resources for further examination.}

Near-term development of highly capable autonomous systems will continue to create new challenges and questions regarding authorization and authentication for AI agents. 

\textbf{Section 3 explores the risks and challenges that could arise from these advanced AI systems, the role different identity and authentication approaches could play in mitigating risks, and the challenges that may arise for these systems.}

\clearpage
\section{Immediate Solutions to Current Use Cases}

\textit{As AI agents are rapidly developed and rolled out, careful thought should be put into authorizing their activities and providing accountability, determining who they act on behalf of, how they are authenticated, monitored, audited, governed, and what scope and permissions these agents are granted. In general, this section builds upon existing, widely accepted specifications for present-day AI agent implementations.} 

\subsection{Agents and their Resources}

At its core, an AI agent interacting with external services, data sources, or tools is acting as a client application. Whether an agent is a sophisticated language model orchestrating a complex workflow or a simpler automated process, its requests to access or manipulate resources beyond its own operational boundaries are analogous to those made by traditional software clients. Consequently, the fundamental principles of authentication and authorization for any client application are equally critical for AI agents. The wealth of knowledge and established best practices developed over years of implementing OAuth, OpenID Connect, and other authentication technologies across web and mobile applications provides a robust starting point for securing agent-based systems.

AI agents may seek access to a diverse array of resources. These can include structured data via APIs (e.g., for customer relationship management, inventory systems, or financial data), unstructured information from knowledge bases or document stores, computational services, or even other AI models. Agent frameworks are extremely varied, and the mechanisms by which agents interact with resources are heterogeneous. Grounding such a discussion can be hard, since agents can, at times, act as both clients and servers. For general purposes, consider an agent as a client workload (synchronous or asynchronous with the user) that makes requests to remote servers. These servers must reliably identify the agent (and/or the user on whose behalf the agent is operating) and determine its permitted actions. 

\subsection{Agent Protocols}

Increasingly, particularly in the context of Large Language Models (LLMs), agents utilize specifically defined ``tools'' or ``plugins.'' These tools are often, although not always,  wrappers around existing REST APIs, equipped with descriptions of their capabilities and input parameters, designed to be discoverable and invocable by an AI model to perform specific actions like sending an email, querying a database, or fetching real-time information.

The increasing sophistication and autonomy of AI agents have spurred the development of specialized communication protocols to standardize how agents interact with remote services or other agents. While many exist, the \href{https://modelcontextprotocol.io/}{Model Context Protocol (MCP)} appears to be leading the pack in adoption, with other protocols such as the \href{https://a2aprotocol.ai/}{Agent-to-Agent Protocol (A2A)} also having wide commercial investment. In essence, MCP is designed to facilitate the connection between model-based interfaces and a diverse ecosystem of external tools and data resources. The evolution of MCP itself underscores the importance of robust authorization discussions: its initial \href{https://github.com/modelcontextprotocol/modelcontextprotocol/pull/133}{design did not include authentication}, but subsequent community feedback and technical discussions have led to the active integration of authentication and authorization considerations. 

\subsection{MCP}

MCP uses a client-server architecture to provide AI models with resources, prompts, and tools. Resources are application-controlled, read-only data sources like files or API responses that give context to the model. In contrast, tools are model-controlled functions that allow the AI to perform actions, such as calling an external API or running a computation. AI applications (clients) connect to MCP servers to access these components. Other features are under consideration, such as \href{https://modelcontextprotocol.io/docs/concepts/elicitation}{elicitation} to request input from users of agents and \href{https://github.com/idosal/mcp-ui}{dynamic UIs} for communication with the user. Communication uses transports like Streamable HTTP or stdio, which support asynchronous operations by allowing the server to push updates to the client.

Given its current trajectory, community engagement, and the illustrative nature of its development, much of the detailed exploration in this section will center on MCP. However, the principles and challenges discussed are broadly applicable to the wider sphere of AI agent protocols and their secure integration with external resources.

\subsection{Authentication} 

Robust authentication is a critical prerequisite for authorization, forming the foundation for secure access to resources by AI agents. The central question for any service is one of authorization: \textit{is this agent permitted to perform this action on behalf of this entity at this time?} Authentication provides the verifiable ``who'' needed to answer that question. When an agent acts on behalf of a user, two distinct authentication challenges must be addressed:

\begin{enumerate}
\item \textbf{Agent Authentication:}The agent software itself must be authenticated as a trusted client. This confirms the agent is the legitimate entity it claims to be, often but not always established through a workload identifier.
\item \textbf{User Authentication \& Delegation:} The human user must be authenticated, and their intent to delegate specific permissions to the agent must be captured.
\end{enumerate}

The ongoing evolution of MPC underscores the importance of \href{https://aaronparecki.com/2025/04/03/15/oauth-for-model-context-protocol}{getting this right}. The community has converged on using OAuth 2.1 as the standard framework, mandating modern security practices like PKCE (Proof Key for Code Exchange) \cite{rfc7636} to prevent authorization code interception attacks.

A key architectural recommendation is that resource servers, such as those implementing MCP, should externalize authentication and authorization decisions to a dedicated authorization server or Identity Provider (IdP) rather than implementing proprietary logic. This separation of concerns is a foundational best practice in modern security architecture and is recommended in the \href{https://modelcontextprotocol.io/specification/2025-06-18/basic/authorization}{MCP specification}. It is particularly effective within a single trust domain, such as a corporate environment. It allows for centralized policy and identity management, enabling IT administrators to use existing corporate credentials via Single Sign-On (SSO) to govern agent configurations and permissions. This approach also benefits end-users by centralizing consent management; instead of approving permissions for each individual tool an agent might use, access can be governed by broader policies, reducing user friction and the risk of consent fatigue.

A significant challenge not yet fully standardized by MCP is how the MCP server authenticates to downstream platforms on the agent's behalf. After the agent authenticates to the MCP server, that server must then make its own authenticated API request to the final tool (e.g., Salesforce, GitHub). Currently, this often relies on custom implementations. 

Finally, maintaining secure agent sessions requires attention to the full authentication lifecycle, especially for asynchronous operations. Long-running tasks may outlive initial access tokens, necessitating secure token refresh strategies that don't compromise the principle of least privilege. This includes implementing proper token expiration or revocation, maintaining audit logs of all authentication events (potentially with identity binding to specific authorizations), and ensuring agent credentials can be promptly revoked when compromised or no longer needed.

\subsection{Dynamic Client Registration} 

The MCP protocol's approach to scalability leveraged Dynamic Client Registration \cite{rfc7591}, allowing any client to register with a server and obtain credentials. While this model offers frictionless onboarding in a ``many-to-many'' ecosystem, it introduces a critical security flaw: it creates a large number of anonymous clients. An unauthenticated, public registration endpoint allows clients to be created without any link to a real developer, organization, or accountable party. This results in a complete lack of a paper trail, opens the door for endpoint abuse (e.g., Denial of Service attacks via mass registration), and makes robust client identification and attestation impossible. For any enterprise or high-security context, this is a high risk.

Various approaches have been proposed to link clients to more robust identities. For example, Client ID Metadata \cite{client_id_metadata}, which proposes supporting URL-based OAuth Client ID Metadata Documents for MCP, letting clients host metadata at HTTPS URLs. Servers fetch and validate it to establish trust without pre-registration or dynamic registration. Client ID metadata could also be used for non-MCP workloads, as could other alternative workload identity paradigms.

\subsection{Authorization} 

Following successful authentication, authorization dictates the specific actions an AI agent is permitted to undertake. It's important to note that, as defined in the MCP specification today, authorization governs the relationship between the MCP client and the MCP server. The mechanism by which an MCP server gains authorization to access downstream APIs or resources on behalf of a user is distinct. 

Within the client-server interaction, authorization leverages standard access control models, such as Role-Based Access Control (RBAC), Attribute-Based Access Control (ABAC), or more fine-grained methodologies. Given the typically non-deterministic nature of language models, least privilege is especially critical when deploying AI agents \cite{delegated_authorization}.

\subsection{Asynchronous Authorization for Agents}

The asynchronous nature of many AI agent workflows creates fundamental challenges for gaining user approval for actions that were not covered in an initial authorization grant. When an agent operates autonomously--potentially executing tasks hours or days after initial user instruction--requiring real-time authorization for sensitive operations becomes impractical. Client Initiated Backchannel Authentication (CIBA) \cite{ciba_spec} provides a solution by decoupling the authorization request from the user's authentication response. CIBA enables a Client to initiate the authentication of an end-user through out-of-band mechanisms, allowing agents to request authorization and continue processing while awaiting user approval. 

This architecture is particularly well-suited to AI agents for several reasons: agents can request authorization for high-risk operations without blocking their entire workflow; users receive notifications on their authentication devices and can approve or deny requests at their convenience; and the system maintains a clear audit trail of all authorization decisions. CIBA supports three delivery modes--poll, ping, and push--each optimized for different agent architectures. In poll mode, agents periodically check for authorization status, which is ideal for batch processing scenarios. Ping mode involves the authorization server notifying the agent when a decision is available, reducing unnecessary network traffic. Push mode delivers the full authorization result directly to the agent, enabling the fastest possible response times. For AI agents operating under regulatory frameworks requiring human-in-the-loop oversight, CIBA provides a standardized mechanism to ensure meaningful human control without degrading the user experience. The protocol's support for binding messages allows agents to provide rich context about requested actions, helping users make informed authorization decisions even when separated from the original task initiation by significant time intervals.

Building on these out-of-band principles, the Model Context Protocol (MCP) also supports an ``elicitation'' capability, which can be extended with a ``URL mode'' (\href{https://github.com/modelcontextprotocol/modelcontextprotocol/issues/1036}{SEP-1036}). This mechanism allows an MCP server to direct a user to an external, trusted URL via their browser for sensitive interactions that must not pass through the MCP client itself. This is particularly relevant for obtaining third-party OAuth authorizations, securely collecting credentials, or handling payment flows, without exposing sensitive data to intermediary systems. By leveraging standard web security patterns and clear trust boundaries, URL mode elicitation provides a secure conduit for obtaining explicit user consent or credentials that can then be used by the MCP server to complete delegated actions. This ensures that even highly sensitive or cross-domain authorizations can be handled securely and asynchronously, maintaining human oversight where required.

\subsection{Identity for AI Agents}

The concept of identity in the context of AI agents is multifaceted and extends beyond simple user impersonation, playing a critical role in authentication, authorization, and auditability.

In current practice using MCP, the host (e.g., Claude Desktop or coding IDEs like Cursor) that the user interacts with connects to an MCP Client to manage communication with a specific MCP server (note that the host and MCP Client may have different identifiers). OAuth 2.1 with PKCE is used to secure the authentication flow for the agent to the MCP server. While this MCP client comes with a client ID generated during dynamic client registration, this is not a robust stand-alone workload identity. While Client ID Metadata could provide a somewhat more robust identifier for the system user, the MCP server (e.g., Claude Desktop is using the MCP), this is still insufficient as a workload identity or AI agent identifier. 

Other workload identity standards could be used here to create robust identifiers for AI agents. For example, the Secure Production Identity Framework for Everyone (SPIFFE) and its runtime environment, SPIRE, provide a model. SPIFFE offers a standardized identity format and a mechanism for issuing cryptographically verifiable identity documents (SVIDs) to workloads automatically, regardless of where they are running. By integrating with SPIRE, an AI agent could be provisioned with a short-lived, automatically rotated identity that it can use to mutually authenticate with other services, establishing trust without relying on static, shared secrets like API keys.

While powerful for establishing a verifiable identity within a controlled infrastructure, the SPIFFE/SPIRE model fundamentally relies on knowledge and control of that infrastructure to attest to a workload's identity. This creates a significant challenge for AI agents, which are designed to operate across heterogeneous trust boundaries where such infrastructure-level trust is not shared. An agent's identity must be portable and verifiable to a third party that has no visibility into its host environment.

However, even robust workload identity frameworks like SPIFFE/SPIRE may not fully address the unique demands of agentic systems at scale. A traditional workload's identity confirms what it is, but in the context of agents, its behaviour is also important. Agent identity must be enriched with metadata about its underlying model, version, and capabilities to enable risk-based access control. Furthermore, because agents are designed to operate across organizational boundaries and act on behalf of users, their identities must be highly portable and natively designed to integrate with the delegated authorization models of OAuth 2.1. This need for a more sophisticated, governable, and interoperable identity for dynamic actors is precisely the gap that emerging agent-specific identity solutions aim to fill.

The growing need for a more governable and feature-rich identity model is driving a significant shift in the commercial Identity and Access Management (IAM) market. Recognizing that the traditional service account is insufficient for the dynamic lifecycle and unique governance needs of AI agents, identity vendors have started to treat them as first-class entities. In addition to identities assigned during client registration, vendors are developing and rolling out dedicated identity management solutions for AI (e.g., Microsoft Entra Agent ID, Okta AIM, and many others). These platforms aim to support the discovery, approval, and auditing of agents using workflows similar to those for human users. These approaches share similarities with traditional IAM service accounts but are purpose-built to meet the need for agents to interact autonomously across trust boundaries. The interoperability between these agent identity systems remains limited, with vendors developing proprietary approaches unless they converge to common standards.

A significant portion of agent activity involves acting on behalf of a human user. Typical workflows do not target this currently, and we will address supporting these on-behalf-of paradigms in the more future-looking Section 3 on delegated identity.

\subsection{SSO \& Provisioning}

Enterprise deployment of AI agents requires a robust identity management infrastructure that integrates with existing corporate systems. Single Sign-On (SSO) through federated identity providers enables users to access agent platforms and administrative interfaces using their existing corporate credentials. This established pattern for human identity must be extended to non-human, agentic identities.

The System for Cross-domain Identity Management (SCIM) \cite{rfc7644} protocol is the standard for automating user lifecycle management, synchronizing access rights with HR systems to grant, update, or revoke permissions as employees join, move roles, or leave an organization. This same lifecycle management is equally critical for the agents themselves, which require formal processes for creation, permissioning, and eventual decommissioning. To address this, experimental work is underway to extend the SCIM protocol to support agentic identities formally. For instance, the proposed System for Cross-domain Identity Management: Agentic Identity Schema Draft \cite{scim_agent_schema} defines a new \textit{AgenticIdentity} resource type. This allows an agent to be treated as a first-class entity within an IAM system, with its own attributes, owners, and group memberships.

By using an extended SCIM schema, organizations can provision agents into services just as they do users. This enables centralized IT administration, where agent permissions are not managed through ad-hoc processes but are governed by the same automated, policy-driven workflows used for human employees. As AI agents proliferate and require access to multiple enterprise applications, managing their lifecycle and consent flows centrally becomes critical for maintaining security, compliance, and operational efficiency. Leveraging a standardized protocol like SCIM for this purpose allows enterprises to apply familiar governance models to this new class of identities.

Crucially, this lifecycle management must extend to a robust and verifiable de-provisioning process. The ``off-boarding'' of an agent is a critical security function, representing the final, authoritative step in its lifecycle. When an agent is decommissioned--or more urgently, when its identity is suspected of compromise--it is not enough to simply revoke its current credentials. A formal de-provisioning signal, such as a SCIM DELETE operation on the agent's \textit{AgenticIdentity} resource, ensures that the identity itself is permanently removed. This action must propagate across all integrated systems, guaranteeing that the agent and all its associated entitlements are irrevocably purged, thereby neutralizing it as a potential persistent threat vector.

\subsection{Operationalizing Agent Identity and Authorization}

The principles of agent identity and authorization are only effective when they can be reliably enforced. A well-established architectural pattern for this is the externalization of authorization logic, which separates the Policy Enforcement Point (PEP) from the Policy Decision Point (PDP) (see NIST SP 800-162 \cite{nist_sp_800_162}). The PEP is the component that intercepts an incoming request (e.g., an API gateway, service mesh sidecar, or middleware), while the PDP is the dedicated service that makes the authorization decision (e.g., ``permit'' or ``deny'') based on defined policies. This separation of concerns allows application developers to focus on business logic while security and platform teams manage policy centrally.

This model is critical in an agent ecosystem, which introduces a proliferation of autonomous decision-making. While an agent itself is a decision point for its own operational logic, the infrastructure it interacts with requires a separate authorization decision point to govern its actions. These architectural patterns provide a concrete location to implement agent-specific security models. The PEP is the ideal place to parse a delegated credential and differentiate between the user who granted authority and the agent acting on their behalf.

Furthermore, when interfacing with existing systems that are not agent-aware, a centralized PEP, like a gateway, becomes a powerful translation layer. It can inspect a modern, rich agent identity token and exchange it for a simple API key or legacy credential expected by a downstream service, providing a crucial bridge between the emerging agent ecosystem and an organization's existing technology investments. Efforts to standardize the communication protocol between a PEP and a PDP are underway in the OpenID Foundation's AuthZEN (Authorization Services) Working Group, which is developing an interoperable API for exactly these types of externalized authorization decisions.

\subsection{Closing the Auditability Gap}

A key benefit of these architectural patterns is closing the critical auditability gap that plagues many current AI systems. Today, an API call made by an agent on a user's behalf is often logged indistinguishably from an action taken directly by the user, creating a black hole for accountability and forensics. By implementing true delegated authority, the credential presented to the PEP contains distinct identifiers for both the human principal and the agent actor. This enables the PEP to generate enriched audit logs that unambiguously record not only who authorized an action but also which specific agent instance performed it. Capturing this rich contextual data is foundational for debugging, meeting compliance requirements, and ultimately building trustworthy autonomous systems where every action can be traced to its origin. For example, within a JWT, \href{https://datatracker.ietf.org/doc/html/rfc8693#section-4.1}{the act (actor) claim} can provide a means to express that delegation has occurred and identify the acting party to whom authority has been delegated.

\subsection{Applying Guardrails}

Guardrails in AI refer to mechanisms, policies, or constraints designed to ensure that AI systems operate safely, ethically, and within intended boundaries. Guardrails can include technical controls, such as limiting access to sensitive data, enforcing usage policies, or monitoring outputs for harmful content. They help prevent unintended behaviors, reduce risks, and maintain trust by guiding AI agents to act responsibly and in alignment with human values.

These mechanisms are a critical extension of the principles found in traditional Identity Governance and Administration (IGA). While a mature IGA program establishes who can access what resources, AI guardrails provide a more specialized, real-time layer of control focused on how an agent uses that access, particularly when data is being exchanged with an AI model. For instance, while IGA may grant an agent permission to access a customer database, an AI guardrail would enforce policies at the point of action, such as automatically masking Personally Identifiable Information (PII) before it is sent to the LLM for summarization. This addresses the unique risks of the agentic paradigm, such as sensitive data leakage to the model and the non-deterministic nature of its outputs. By integrating AI-specific guardrails with a strong IGA foundation, organizations can create a defense-in-depth strategy that governs not just an agent's entitlements, but its behavior.

\subsection{Agent to Agent}

MCP is a common paradigm for accessing resources. In some cases, a tool call to a remote MCP server is being used to request an action or response from an external AI agent. A broader example of this is the new agent-to-agent (A2A) protocol, designed to allow agents to engage in structured communication with other agents. Many other similar protocols exist, all with the vision to enable interagent communication and task completion. A2A provides an \href{https://a2a-protocol.org/dev/specification/#4-authentication-and-authorization}{outline} of how authentication should be performed, but leaves many questions highlighted above unanswered. A2A introduces additional complexities when authorization extends beyond access to another agent and into setting restrictions on the scope of actions or resource use for downstream agents. We will explore this more in section three.

\subsection{Summary of Immediate Solutions}

\textbf{Agents and today's authentication and authorization patterns can work for \textit{synchronous} agents using multiple tools across a \textit{single trust domain,} but not asynchronous or multi-domain contexts.}

Today's authentication and authorization solutions for AI agents provide an effective and well-understood pattern for the foundational use case: a single agent accessing multiple tools within a unified trust domain. When an enterprise user interacts with an AI assistant that needs to query their company's CRM, update a project management tool, and fetch data from an internal knowledge base, existing OAuth 2.1 flows with PKCE, combined with protocols like MCP, provide robust security. These scenarios benefit from a shared identity provider, consistent authorization policies, and centralized consent management--the agent receives a workload identity, authenticates via the corporate IdP, and accesses various internal tools using scoped permissions managed by IT administrators. However, this well-solved pattern represents just the tip of the iceberg. The moment agents need to operate across trust boundaries--such as an enterprise agent accessing both internal Salesforce data and external market research APIs, or when agents begin delegating tasks to other agents that may reside in different security domains--current frameworks reveal significant gaps. 

This model begins to break down, in part because identity mechanisms rooted in control over a specific infrastructure (such as SPIFFE/SPIRE) do not naturally extend across organizations that do not share visibility or control over infrastructure. As shown in the next section, the challenges multiply exponentially with recursive delegation (agents spawning sub-agents), scope attenuation across delegation chains, true on-behalf-of user flows that maintain accountability, and the interoperability nightmare of different agent identity systems attempting to communicate. While it is possible to securely connect one agent to many tools within a single organization's control, the broader vision of autonomous agents seamlessly operating across the open web remains largely unsolved, the subject of the next section.

\begin{figure}[h!]
    \centering
    \includegraphics[width=0.7\textwidth]{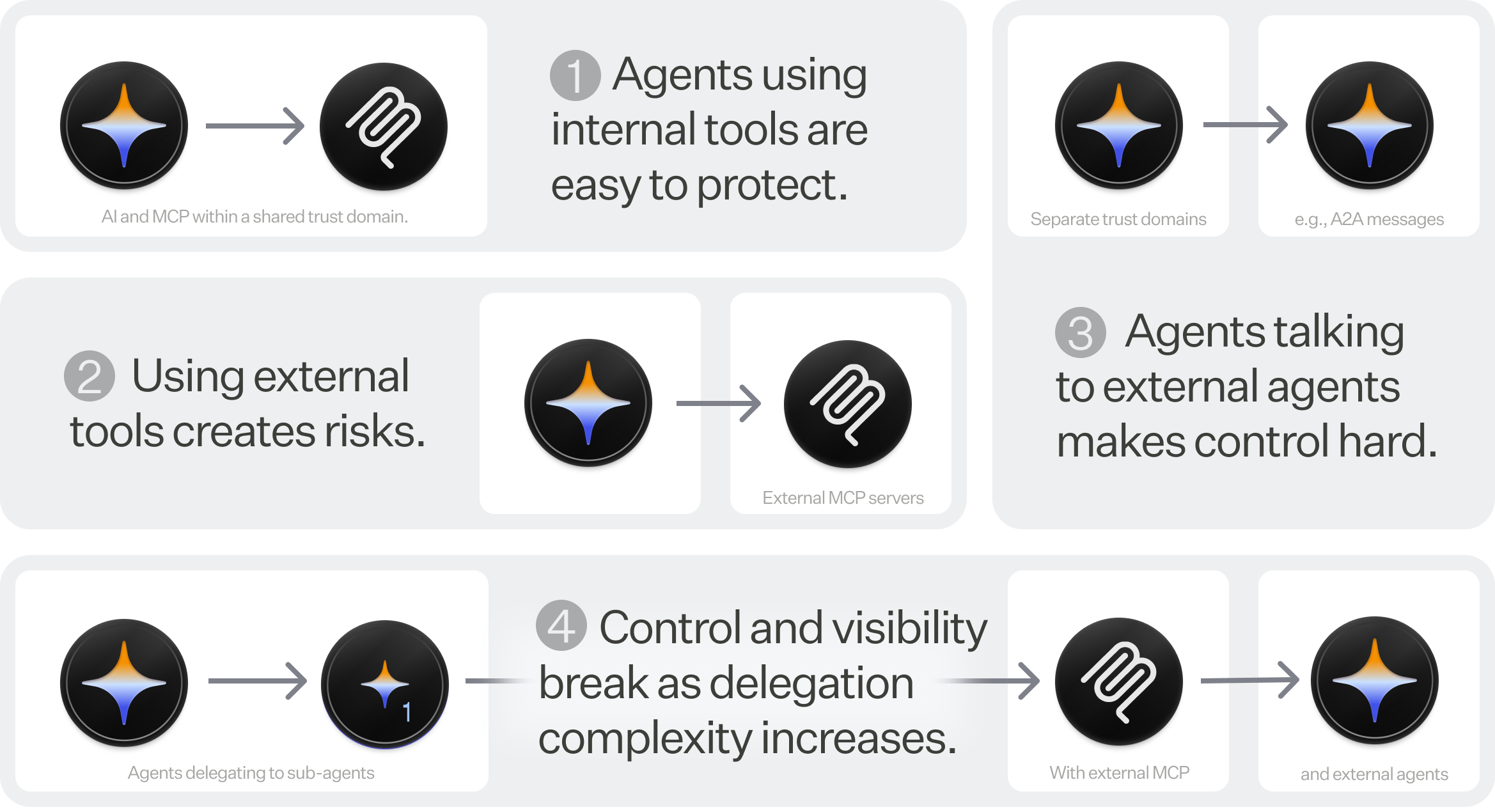}
    \caption{An illustration of the types of agent to tool (MCP) and agent to agent (A2A) within trust domains (e.g., a single enterprise) and across trust boundaries. Complexity, security risks, observability, and identity challenges all become harder as external communication and delegation increase.}
    \label{fig:agent-to-agent}
\end{figure}

\begin{minipage}{\textwidth}
\fbox{
\begin{minipage}{0.95\textwidth}
\vspace{0.5em}
\textbf{Best Practices} 

Several general-purpose best practices emerge from the convergence of AI agents, MCP, and enterprise identity management:
\\

\begin{itemize}
    \item \textbf{Use standard protocols}: Implement open frameworks like OAuth 2.1 \cite{oauth_v2_1} for authentication and SCIM for lifecycle management rather than custom authentication, authorization, and provisioning mechanisms.
    
    \item \textbf{Authenticate agent interactions}: Most agent-to-resource and agent-to-agent interactions should be authenticated. While some contexts don't require authentication, high-security contexts should never allow anonymous access.
    
    \item \textbf{Apply least privilege rigorously}: Agents should not be granted broad access to protected resources, and their permissions should be managed in a way that is consistent with the permissions of the users to whom they will output information.
    
    \item \textbf{Automate agent lifecycle management}: Address complex events like agent provisioning and de-provisioning across multiple systems, and ownership transfer by profiling standards like SCIM, rather than tightly coupling agent management to the workflow of a single user.
    
    \item \textbf{Maintain clear audit trails for governance}: Log all authentication events, authorization decisions, and agent actions. This enables meeting future compliance requirements, discourages malicious agent actions, and facilitates forensic detection of fraud.
    
    \item \textbf{Design for interoperability}: Build adaptable identity systems that can evolve with emerging standards like IPSIE and future MCP specifications.
\end{itemize}
\vspace{0.5em}
\end{minipage}
}
\end{minipage}

\clearpage
\section{Future-looking Problems for Autonomous Agent Identity and Authorization} 

\textit{While the solutions in Section 2 address current needs, the trajectory of AI development points toward agents operating at a far greater scale and with higher degrees of autonomy. This leap forward introduces a new class of complex, future-looking challenges for identity and access management. These problems move beyond simple client-server authentication and demand a fundamental rethinking of identity, delegation, consent, and governance in a world populated by millions of non-human actors.}

\subsection{Architectural Models for Agent Identity}

The most fundamental challenge is establishing who or what an agent is. Today, an agent's identity is often just a client ID, which is uninformative, undifferentiated for traditional workloads, and insufficient for a scalable, secure ecosystem. As agent workloads proliferate, robustly and interoperably identifying them becomes paramount. Beyond an identifier, an agent can possess metadata or attributes that describe its nature, capabilities, discovery, and governance. These attributes can also influence its entitlements. 

A standard for agent identity is only effective if it supports the architectural patterns both enterprises and consumers need. Fragmentation is already a risk, with vendors developing proprietary systems. To avoid a future where agents require dozens of identities to operate, a few key models exist that are worth considering:

\begin{itemize}
\item \textbf{The Enhanced Service Account}: The most likely near-term enterprise pattern is that this model extends the familiar concept of workload identity. An agent is treated like a service, but its identity token is enriched with agent-specific metadata (e.g., agent\_model, agent\_provider, agent\_version), asserted via standards like SPIFFE/SPIRE or proprietary extensions.  
\item \textbf{The Delegated User Sub-Identity}: Foundational for agents acting directly on behalf of a user, this model creates an identity that is intrinsically linked to and derived from that user's session. It is the formal implementation of the ``on-behalf-of'' (OBO) flow, where the agent's identity is distinct but inseparable from the user's authority.  
\item \textbf{Federated Trust and Interoperability}: Agents require an interoperable trust fabric to operate across diverse domains without a central IdP. This fabric can be built using established frameworks, such as OpenID Federation (with HTTPS-based identifiers), or systems that leverage X.509 certificates, enabling verification between different OpenID and non-OpenID identity systems. This is a rich and emerging body of work.   
\item \textbf{Sovereign and Portable Agent Identity}: Each agent instance can be assigned a globally unique and verifiable identifier for accountability, using schemes like DIDs or others currently being standardized. By managing its own cryptographic keys, the agent can directly assert its identity in peer-to-peer interactions, enabling a more open and decentralized ecosystem.
\end{itemize}

Proposals such as OpenID Connect for Agents (OIDC-A) \cite{openid_connect_agents} aim to standardize this by defining core identity claims, capabilities, and discovery mechanisms. Such standards are crucial for ensuring interoperability and providing the granular identification needed for robust audit trails and safety. More generally, identifying agents \cite{identifying_agents} will involve knowing the specific instance that took an action and the properties of that system.

\subsection{Delegated Authorization and Transitive Trust}

Once an agent has an identity, it needs the authority to act. Currently, agents often \textbf{impersonate} users in a manner that is opaque to external services (e.g., via screen scraping and browser use), creating significant accountability gaps and security risks. The solution is to move to a model of explicit \textbf{delegated authorization} \cite{delegated_authorization}.

\subsubsection*{From Impersonation to Delegation (On-Behalf-Of)}

The OBO pattern is not a new problem, but the proliferation of AI agents has made it a critical challenge to solve at scale. This has led to more prescriptive guidance on how to implement dynamic delegation and consent using existing standards. The foundational pattern is a true OBO flow, as explored in proposals like ``OAuth for AI Agents on Behalf of Users'' \cite{oauth_ai_agents_obo}. This is critically different from impersonation because it results in an access token containing two distinct identities: the user who delegated authority (e.g., in the \textit{sub} claim) and the agent authorized to act (e.g., in the \textit{act} or \textit{azp} claim). This creates a clear, auditable link from the very first step.

\subsubsection*{Recursive Delegation and Scope Attenuation}

The true power of an agent ecosystem emerges from recursive delegation: the ability of one agent to decompose a complex task by delegating sub-tasks to other, more specialized agents. This is a foundational pattern for building sophisticated applications, allowing a primary agent to orchestrate a network of agents to achieve a goal. This modularity, however, introduces the immense security challenge of managing authorization across a multi-hop delegation chain. This creates a significant transitive trust problem, as a resource server at the end of the chain must be able to cryptographically verify the entire delegation path back to the original user, not just the final sub-agent making the request. This end-to-end verifiability is foundational for creating an unambiguous audit trail that links every action to its origin. The challenge is magnified when the chain crosses trust domain boundaries, as an agent from one organization delegates to another, requiring a robust identity federation model to preserve the authorization context across disparate security systems.

Solving this requires \textbf{scope attenuation}: the ability to progressively and verifiably narrow permissions at each step in the delegation chain. The choice of mechanism depends on the required trust model and architecture. For centrally-managed or hub-and-spoke ecosystems, OAuth 2.0 Token Exchange \cite{rfc8693} provides a standardized, online approach where an agent requests a down-scoped token from an authorization server on behalf of a sub-agent. This centralizes policy control and simplifies revocation, but introduces latency. In contrast, for more decentralized and dynamic agent networks, modern capability-based token formats like Biscuits \cite{biscuit_sec} and Macaroons \cite{macaroons} enable offline attenuation. These tokens allow a holder to create a more restricted version of a token without contacting the original issuer, embedding authority and constraints within the credential itself. This approach is underpinned by the Object-Capability (OCap) security model, where possession of the token is proof of authority. While OCap offers powerful, fine-grained security, its integration with existing web standards and the challenge of offline revocation remain areas of active exploration.

\subsubsection*{The Revocation Challenge}

A critical, and largely unsolved, problem in these architectures is \textbf{revocation}. With traditional OAuth 2.0, revoking a bearer token can be challenging. In a decentralized system using offline-attenuated tokens, the problem is magnified. If a user revokes the primary agent's access, there is no clear, immediate mechanism to propagate that revocation down a chain of offline tokens that may have already been further delegated.

Several standards-based approaches are converging to solve this. The OpenID Foundation's Shared Signals Framework defines a protocol for communicating security events, allowing for the near-real-time propagation of revocations. Ensuring these mechanisms are implemented consistently is a key goal of enterprise profiles like those being developed by the Interoperability Profiling for Secure Identity in the Enterprise (\href{https://openid.net/wg/ipsie/}{IPSIE}) working group, which includes requirements for the reliable termination of sessions. One emerging mechanism for this is OpenID Provider Commands, a protocol that enables an OpenID Provider (OP) to send direct, verifiable commands--such as ``Unauthorize''--to a Relying Party (RP) to terminate a specific user account's session. By leveraging such standards, a user's decision to revoke an agent's access at the identity provider can be reliably propagated throughout the ecosystem.

This inability to guarantee timely, system-wide revocation makes proactive risk mitigation essential. Instead of relying solely on time-based expiration, which is ill-suited for high-velocity agents, credentials can be constrained by execution counts. This approach grants an agent a strictly limited number of operations, ensuring that even if revocation is delayed, the potential impact is predictably bounded. This technique is a powerful tool for enforcing least privilege and managing trust in these complex, autonomous systems.

\subsubsection*{De-provisioning \& Off-boarding}

While revocation addresses the immediate termination of an agent's active session, de-provisioning represents the permanent and complete removal of the agent's identity and its associated entitlements. The distinction is critical; a compromised agent identity that is merely ``revoked'' may retain its underlying registration and trust relationships, representing a dormant but persistent threat. De-provisioning is the ultimate response to a compromise or end-of-life event, but its implementation differs substantially between enterprise and consumer contexts.

In an enterprise, an agent is treated as a non-human employee, and its off-boarding must be a structured, verifiable process. When triggered by an event like a security alert, the agent's core identity is terminated in the central IdP (ideally via a SCIM DELETE operation), which also invalidates all associated credentials. The IdP must then broadcast a de-provisioning signal to all federated domains using protocols like the Shared Signals Framework (SSF). This ensures the agent's identifier is purged from all access control lists (ACLs) to prevent orphaned privileges, and any stateful resources it owns are securely transferred or decommissioned. Failure in this process creates a persistent backdoor, leaves behind exploitable data assets, and can result in severe audit and compliance failures.

For consumer platforms, de-provisioning is a matter of upholding user trust and privacy. The process is initiated directly by the user, such as when they delete a custom agent or revoke its platform permissions.

Ultimately, while sharing principles with traditional identity lifecycle management, the de-provisioning of AI agents represents a fundamentally distinct and more critical challenge. Unlike human identities, whose lifecycles are slow and centrally managed, or traditional workloads, which are often confined to a predictable scope, agents are designed for high-velocity, autonomous, and cross-domain operation. This combination makes them uniquely dangerous if compromised. An agent wields the delegated authority of a human but operates with the speed and scale of a machine, creating a vastly amplified blast radius for a potential breach. Furthermore, its ability to recursively delegate authority means a single compromised identity can trigger a cascading failure across an entire ecosystem of sub-agents. Consequently, robust de-provisioning is not merely an operational best practice; it is a foundational pillar of safety and trust. Without the verifiable, high-speed capability to permanently erase a rogue agent's existence across all trust boundaries, we cannot build a secure and governable autonomous ecosystem.

\subsection{Registries and Dynamic Connections to External Tools}

An exciting capability of autonomous agents is their ability to dynamically discover and connect to new tools and services based on user intent. This necessitates a robust infrastructure for service discovery. A prime example of this is \href{https://github.com/modelcontextprotocol/registry}{MCP Registries}, an open catalog designed to standardize how MCP servers are published and discovered by clients. While such registries solve the critical problem of discoverability, they introduce a significant challenge for identity and access management: establishing trust on first contact. When an agent queries a registry and identifies a new, previously unknown server to fulfill a task, the resource server has no pre-existing trust relationship with the agent, and conversely, the agent's user has no basis to trust the server. 

This raises a critical question for user experience and security, particularly for semi-autonomous agents operating in the background: how does the user grant authentication and delegate permission for an interaction with a service they may not have known existed moments before? This scenario moves beyond simple, upfront consent and necessitates architectures that can handle asynchronous, out-of-band authorization. Frameworks such as Client Initiated Backchannel Authentication (CIBA), discussed earlier, become essential, providing a mechanism for the agent to pause its workflow and securely request explicit user approval on a trusted device before establishing a connection and sharing any data. This dynamic trust establishment is a foundational requirement for building an open and interoperable agent ecosystem, rather than one confined to pre-approved, walled gardens of services. These challenges with MCP registries similarly map to agent-to-agent style communications or external requests made in natural language to third parties.

\subsection{Scalable Human Governance and Consent}

As agents proliferate, the sheer volume of their actions creates a fundamental scalability challenge for human oversight. Regulatory frameworks like the EU AI Act \cite{eu_ai_act} mandate ``effective oversight'' for high-risk AI (Article 14), but requiring human approval for every autonomous action is impossible. A single user could have dozens of agents making thousands of daily decisions, leading to an unmanageable deluge of permission prompts. This \textbf{`consent fatigue'} not only degrades user experience but paradoxically reduces security as users begin reflexively approving requests.

\subsubsection*{Designing for Scalable Governance}

Addressing this requires moving beyond traditional interactive consent to new architectural patterns for governance:

\begin{itemize}
\item \textbf{Policy-as-Code for Agent Authorization.} Instead of users clicking ``approve'' for every action, an administrator or user defines a high-level policy that sets the agent's operational envelope (e.g., budgetary limits, data access tiers, API call velocity). The IAM system then enforces this policy programmatically.  
\item \textbf{Intent-Based Authorization.} Users approve a high-level intent in natural language (e.g., ``Book my travel for the upcoming conference''). The system translates this into a bundle of specific, least-privilege permissions, which are then enforced under the hood.  
\item \textbf{Risk-Based Dynamic Authorization.} A policy decision point can assess the risk of an agent's requested action in real-time. Routine, low-risk actions are permitted automatically. However, an anomalous request would dynamically trigger a \textbf{Client Initiated Backchannel Authentication (CIBA)} flow to request explicit, out-of-band human approval.
\end{itemize}

\subsubsection*{Natural Language Scopes}

Users naturally express intent in plain language (``help me with the report, but don't access confidential data''), which is flexible but lacks the precision needed for security enforcement. The solution lies in a hybrid approach: using AI to help translate high-level natural language instructions into formal, machine-readable access control policies. The user approves the intuitive instruction, while the system enforces auditable, deterministic resource constraints, ensuring the agent remains bounded even if it misinterprets the user's intent.

\subsubsection*{Guardrails as Risk Mitigation}

Guardrails act as a critical layer of defense, preventing undesirable agent behaviors and ensuring adherence to predefined boundaries. They directly address several key problems inherent in autonomous systems:

\begin{itemize}
\item \textbf{Preventing Unintended Information Sharing} - Guardrails can enforce strict access controls, ensuring that only approved data and resources are exchanged with AI models. This prevents data breaches and information leakage while protecting sensitive information.  
\item \textbf{Masking Sensitive Information} - Guardrails can automatically detect and mask sensitive data, such as PII or financial details, before it's processed or shared with an AI model for task execution, further enhancing data privacy and security.  
\item \textbf{Controlling Unintended Actions} - By defining permissible actions and outputs, guardrails can stop agents from performing actions outside their intended scope, even if their core programming has a bug or misinterprets an instruction.  
\item \textbf{Limiting Resource Consumption} - Agents can sometimes consume excessive computational resources or make too many API calls. Guardrails can set rate limits and resource quotas to prevent system overload and unnecessary costs.  
\item \textbf{Maintaining Compliance} - Regulatory frameworks and internal policies often necessitate specific behaviors and restrictions. Guardrails can programmatically enforce these compliance requirements, reducing legal and reputational risks.  
\item \textbf{Ensuring Ethical Alignment} - Guardrails can be designed to prevent agents from generating harmful, biased, or unethical content or from engaging in discriminatory actions.
\end{itemize} 

\subsection{Advanced Challenges and Broader Implications}

Beyond these core issues, several other advanced challenges loom on the horizon.

\subsubsection*{Binding Identity to Action and Output}

Merely identifying an agent is insufficient; we must be able to irrefutably bind that identity to the actions it performs and the content it generates. This is foundational for establishing accountability, non-repudiation, and auditability. Initiatives like the Coalition for Content Provenance and Authenticity (C2PA) \cite{c2pa_spec}, which provides tamper-evident metadata for digital assets, offer valuable lessons for creating verifiable audit trails for agent-driven activities.

\subsubsection*{Privacy vs. Accountability}

Identified agents operating on behalf of users create a deep tension between accountability and privacy. The very traceability needed for audits enables cross-domain tracking that can create comprehensive and potentially sensitive behavioral profiles. \textbf{Selective disclosure} mechanisms--leveraging cryptographic techniques like zero-knowledge proofs and anonymous credentials--offer a path forward. These allow an agent to prove a specific claim (e.g., ``is authorized to access medical data'') without revealing its underlying identity, but integrating these techniques with existing identity standards and regulatory requirements remains a significant challenge.

\subsubsection*{The Presentation-Layer Problem: Browser and Computer Use Agents}

A distinct class of agents, such as OpenAI's Operator, operates by directly manipulating user interfaces (browsers, GUIs) rather than calling APIs. This inverts the security model, as these agents effectively impersonate human users at the presentation layer, bypassing all traditional API-based authorization controls. Differentiating their actions from the user's is nearly impossible, creating a significant gap in our current authorization frameworks, which initiatives like Web Bot Auth are beginning to address by creating identity and authentication mechanisms specifically for this presentation-layer interaction.

\subsubsection*{Protecting the Open Web and Differentiating Agents}

Finally, the proliferation of agents exacerbates the age-old problem of detecting bots online, but with a critical new dimension: the need to distinguish between malicious bots and legitimate, value-adding AI agents. As websites deploy more aggressive bot-blocking to prevent data scraping and abuse \cite{prevent_data_scraping}, well-behaved agents performing tasks on a user's behalf risk being locked out. This creates a pressing need for a standardized way for responsible agents to browse the web.

A promising approach is emerging in the form of \textbf{Web Bot Auth}, a proposal at the IETF \cite{web_bot_auth} that allows an agent to prove its identity directly within its HTTP requests cryptographically. This method acts as a ``passport for agents,'' using HTTP Message Signatures to attach a verifiable identity to traffic, regardless of its IP address. This initiative is gaining significant traction through a collaboration between agent infrastructure providers like Browserbase and major web security and platform companies such as Cloudflare and Vercel. 

Key here is differentiating browser-centric authentication from the workload identity models discussed for API-based agents (e.g., via MCP). Web Bot Auth authenticates the agent platform to the web server to prove it is a responsible actor on the open web. In contrast, workload identity for API agents authenticates the agent to a specific, permissioned API endpoint, often as part of a delegated authorization flow on behalf of a user. The former is about establishing a baseline of trust for public web access, while the latter is about enforcing granular permissions for private resource access.

Robust agent identification via protocols like Web Bot Auth could enable a more nuanced, two-tiered web, where identified, trusted agents are granted permissioned access, while anonymous agents are restricted. This raises profound questions about the future of the open web and the need to differentiate between human and agent traffic, a challenge that extends from technical proofs-of-humanity \cite{proofs_of_humanity} to commercial identity verification services.

\subsection{The Economic Layer: Identity, Payments, and Financial Transactions}

The utility of autonomous agents is, in part, tied to their ability to engage in economic activity, such as accessing paid data sources, purchasing goods, or orchestrating services. This capability introduces a fundamental challenge to existing e-commerce and API security models, which are largely predicated on direct human interaction and consent at the point of transaction. This shift necessitates new protocols to manage authorization, verify user intent, and ensure accountability in agent-driven commerce. Several emerging standards address different facets of this problem, each with distinct mechanisms and intended applications.

\subsubsection*{FAPI: Securing High-Consequence APIs}

For any transaction involving high-stakes and irreversible actions, such as bank transfers or securities trades, the underlying API interactions must adhere to the highest security standards. The FAPI 1.0 and 2.0 specifications from the OpenID Foundation provide an established security profile for this purpose. FAPI is not agent-specific; rather, it hardens the OAuth 2.1 framework to protect high-risk resource servers. Its mandates, including sender-constrained access tokens (via mTLS or DPoP), stronger client authentication, and strict consent-logging requirements, provide a foundational security layer. Any agent system engaging in regulated or high-value financial operations would need to interact with APIs that are protected by this FAPI profile to ensure transactional integrity and non-repudiation.

\subsubsection*{Agent Payments Protocol (AP2): Verifiable Intent for Commercial Transactions}

Operating at a layer above API security, the new \href{https://cloud.google.com/blog/products/ai-machine-learning/announcing-agents-to-payments-ap2-protocol}{Agent Payments Protocol (AP2)} from Google is designed to address the specific problem of capturing and verifying user intent in autonomous commercial transactions. Developed as an extension for protocols like A2A and MCP, its primary mechanism is the Mandate, a cryptographically-signed digital artefact that serves as auditable proof of a user's instructions. AP2 defines a two-stage process to create a non-repudiable audit trail:

\begin{itemize}
\item \textbf{Intent Mandate} - When a user gives a high-level instruction captured as a signed Intent Mandate. This provides the auditable context for the entire interaction.  
\item \textbf{Cart Mandate} - Once the agent finds an action that meets the criteria, the user's approval of that specific purchase signs a Cart Mandate. For pre-authorized tasks, the agent can generate this on the user's behalf if the conditions of the Intent Mandate are precisely met.
\end{itemize}

This chain of evidence, often signed using Verifiable Credentials (VCs) that bind the Mandates to a user's identity, directly answers the critical questions of authorization and authenticity.

\subsubsection*{KYAPay: Identity-Linked Tokens for Programmatic Onboarding}

A distinct challenge arises when an agent must interact with a new service for the first time, where no pre-existing user account or payment relationship exists. The \href{https://www.kyapay.ai/}{KYAPay protocol} addresses this ``cold start'' problem by tightly coupling identity and payment authorization into a single, portable token. The protocol defines a ``Know Your Agent'' (KYA) process, extending traditional KYC/KYB identity verification to the agent itself. The output is a JSON Web Token (JWT) that bundles verified identity claims with payment information. This allows an agent to perform programmatic onboarding and payment in a single, atomic interaction with a new service. 

\subsection{Part 3 Conclusion}

Finally, building a scalable agent ecosystem requires addressing the full operational picture. An agent's identity is not static; it requires robust \textbf{lifecycle management}--from secure creation and registration, through permission updates, to eventual, verifiable decommissioning. This managed identity becomes the anchor for discoverability, enabling agents to find and interact with trusted services through secure, authenticated registries rather than operating in an unvetted wilderness. This entire identity-aware infrastructure, in turn, can serve as a critical \textbf{policy decision point}. The authorization layer becomes the ideal place to implement systemic guardrails, enforcing not just access control but also regulatory constraints, safety protocols, and responsible AI principles. Crucially, for this ecosystem to thrive, it must not become a walled garden. 

\clearpage
\section{Example Use Cases for Robust Agent Authorization} 

\textit{To make the future challenges of agent identity and access management concrete, this section outlines six scenarios ordered by increasing complexity. Each case illustrates a distinct failure mode of traditional Identity and Access Management (IAM) frameworks when confronted with the unique operational characteristics of AI agents, demonstrating the need for new, agent-centric solutions.}

\subsection{High-Velocity Agents and Consent Fatigue}

The most immediate challenge arises from the sheer velocity of agent actions within a single trust domain. Consider an enterprise AI agent tasked with optimizing a digital advertising budget. A high-level command from a marketing analyst: ``Reallocate budget to maximize click-through rate'', could translate into hundreds of discrete API calls to pause campaigns, adjust bids, and transfer funds in mere seconds. A traditional IAM model predicated on synchronous, user-mediated consent for each sensitive action is untenable. The user would face an unmanageable stream of authorization prompts, leading to \textbf{consent fatigue} and the reflexive approval of requests without due diligence. This scenario proves that for high-velocity agents, per-action authorization must be replaced by a more robust model of pre-authorized, \textbf{policy-based controls}. The agent must operate within a clearly defined operational envelope (e.g., budgetary limits, approved targets) enforced by the resource server, shifting the security posture from interactive consent to programmatic governance.

\subsection{Asynchronous Execution and Durable Delegated Authority}

The next level of complexity involves agents executing long-running, asynchronous tasks. An enterprise process agent, for example, might be assigned to onboard a new employee--a workflow spanning days or weeks. The agent must interact with multiple internal services to provision hardware from IT, create an identity in the HR system, and enroll the user in benefits programs. IAM models based on short-lived, user-session-bound access tokens are fundamentally incompatible with this pattern. The agent requires a durable, \textbf{delegated identity} that is a first-class citizen in the IAM system, distinct from the initiating user, allowing it to authenticate independently over extended periods. Furthermore, if a step in the workflow requires an exceptional approval (e.g., a signing bonus exceeding policy limits), the agent must have a standardized mechanism to escalate. Protocols like the \textbf{Client-Initiated Backchannel Authentication (CIBA)} flow provide a solution, enabling the agent to request secure, out-of-band authorization from the appropriate human decision-maker without halting its entire operation.

This durable identity, however, also represents a high-value target. If compromised, the agent becomes a persistent threat vector capable of acting maliciously across multiple enterprise systems for an extended period. This elevated risk profile proves that a simple token revocation is an inadequate response; the scenario demands an immediate and complete cross-system \textbf{de-provisioning} capability to ensure the compromised identity is permanently neutralized across the HR system, IT infrastructure, and all other integrated services.

\subsection{Cross-Domain Federation and Interoperable Trust}

When an agent's tasks cross organizational boundaries, the limitations of siloed, enterprise-specific IAM become apparent. Consider a financial advisory agent that, on behalf of a user, must aggregate data from the user's bank (Trust Domain A), a third-party investment platform (Trust Domain B), and a credit reporting agency (Trust Domain C). In this scenario, no single Identity Provider (IdP) can serve as the source of truth for both identity and authorization across all domains, and the agent faces the critical challenge of proving to Domain B that it has legitimate, user-delegated authority originating from Domain A. The solution requires a federated architecture built on interoperable standards that allow trust to be securely transferred across these boundaries. For complex, multi-hop workflows, the IETF's work on Identity and Authorization Chaining Across Domains \cite{identity_chaining} defines a pattern using OAuth 2.0 Token Exchange to preserve the original identity context. An agent can present a token from one domain to another, which exchanges it for a new token that carries forward the original user and client identity, ensuring a complete audit trail. 

In more centralized enterprise scenarios, the Identity Assertion Authorization Grant \cite{identity_assertion_authz_grant} draft provides another mechanism, allowing an agent to use an identity assertion from a trusted corporate IdP to obtain an access token for a third-party API, enabling centralized control over cross-application access. Alternatively, in more decentralized ecosystems where a common IdP may not exist, the agent can present a verifiable credential that cryptographically encapsulates its delegated authority. Each of these approaches illustrates the necessity of evolving IAM from a centralized function into a standardized, interoperable trust fabric built on the core principle of enabling verifiable, auditable proof of delegated authority that can be securely passed and understood across disparate security domains.

\subsection{Recursive Delegation in Dynamic Agent Networks}

Looking toward future architectures, a primary agent may need to compose tasks by delegating to networks of specialized, third-party agents discovered and engaged in real-time. This introduces the critical challenge of \textbf{recursive delegation}, where an agent must pass a subset of its authority to a sub-agent (similar to a Russian nesting doll). An IAM model for this reality must support multi-hop delegation chains where permissions are progressively narrowed at each step to enforce the principle of least privilege--a process known as \textbf{scope attenuation}. For example, a primary agent with broad data analysis permissions could delegate a specific data collection task to a sub-agent, but grant it only a fraction of its own access rights and operational budget. Trust in such a network is decentralized and must be proven at each step. This likely requires token formats like \textbf{Biscuits} or \textbf{Macaroons}, which allow a token holder to create a more restricted version of a token offline. Authority becomes embedded and verifiable within the credential itself, removing the need for constant callbacks to a central authorization server and enabling secure, decentralized collaboration.

\subsection{IAM as a Safety System for Cyber-Physical Agents}

The ultimate challenge for IAM lies in governing autonomous agents whose actions have direct and potentially irreversible consequences in the world. For an agent managing a city's water distribution network or a fleet of autonomous delivery drones, authorization is no longer about controlling data access; it becomes a fundamental component of the system's safety case. The delegated authority must be expressed as a complex, machine-readable policy that defines a safe operational envelope (e.g., ``maintain reservoir levels between X and Y; never exceed pressure Z''). The agent's identity must be irrefutably bound to its actions to enable forensic analysis and ensure non-repudiation. For high-consequence decisions that fall outside this envelope, the agent must trigger a high-assurance, auditable escalation path to a human operator, ensuring that human judgment is the final arbiter for actions with real-world impact. In these cyber-physical systems, IAM transcends its traditional role and becomes a core safety and policy enforcement layer.

\subsection{Agents Acting on Behalf of Multiple Users}

OAuth was designed to enable workloads to access protected resources on behalf of one human using a subset of that human's permissions. Agents are increasingly being used as part of a team, and the output of an agent may be written into a codebase or chat channel to which multiple users have access. If the agent is only acting on behalf of only one user, it may gather context via MCP or A2A, which other users do not have access to, and it may include that information in its output. For example, you could imagine a CFO having an agent answer questions in a chat channel. While the CFO may have access to salary data, not every member of the channel does. As a result, the agent may disclose confidential salaries because it can act under the CFO's permission. It lacks a standardized method for respecting the subset of permissions that overlap for all users in the channel. Attribute-based access control (ABAC) and fine-grained authorization can work to address this, but complexity and challenges appear in any implementation. 

\clearpage
\section{Conclusion}

The rapid evolution of AI agents from simple tools into autonomous actors marks a critical inflection point for the digital identity landscape. As this paper has outlined, the industry is not starting from scratch. Existing foundational frameworks provide robust and immediately applicable solutions for securing today's agents. The best practices of separating concerns, applying least privilege, and ensuring clear audit trails are the bedrock upon which the next generation of agentic systems must be built.

However, the future of a truly interconnected and autonomous agent ecosystem calls implementers to look beyond this foundation. It invites a pioneering new era of identity and authority defined by \textbf{true delegation over impersonation}, \textbf{scalable governance over consent fatigue}, and \textbf{interoperable trust over proprietary silos}. Solving for recursive delegation, scope attenuation, and verifiable, enterprise-grade security profiles is the central work of this time.

\textbf{This is a call to action.} Successfully navigating this future depends on a concerted, collaborative effort across the industry, and there are clear avenues for contribution:

\begin{itemize}
\item \textbf{For developers and architects,} the immediate task is to build on the secure foundation of existing standards while designing systems with the flexibility to incorporate emerging models of delegated authority and agent-native identity. Crucially, this means aligning with enterprise profiles, like those developed by IPSIE, to ensure their solutions are secure, interoperable, and ready for enterprise adoption.  
\item \textbf{For standards bodies,} the challenge is to accelerate the development of protocols that formalize these new concepts, ensuring that the future ecosystem is built on a foundation of interoperability rather than a patchwork of proprietary, fragmented identity systems.  
\item \textbf{For enterprises,} the imperative is to begin treating agents as first-class citizens within their IAM infrastructure and establish robust lifecycle management, from provisioning to secure, verifiable de-provisioning, governance policies, and clear lines of accountability.
\end{itemize}

The journey from authenticating simple clients to establishing trustworthy identities for autonomous agents is not just a technical upgrade; it is a fundamental evolution in how we manage trust online. By embracing these challenges as opportunities for innovation, we can collectively build an ecosystem where the immense potential of AI agents is unlocked securely, responsibly, and for the benefit of all.

\clearpage
\section*{Key Terms and Acronyms}
\begin{longtable}{p{0.3\textwidth} p{0.65\textwidth}}
    \toprule
    \textbf{Term} & \textbf{Definition} \\
    \midrule
    \endfirsthead
    
    \toprule
    \textbf{Term} & \textbf{Definition} \\
    \midrule
    \endhead
    
    \bottomrule
    \endfoot

Artificial Intelligence (AI) & A system that can intelligently respond to user intent or instructions. Typically backed by a language model.  \\
AI Agent & An AI-based system that can take autonomous `action' based on `decisions' made at model inference time to achieve specific goals. \\
Authentication & The process of verifying an identity. For agents, this includes authenticating both the agent software itself (Client Authentication) and the human user delegating authority. \\
Authorization & The process of determining the specific actions and resources an authenticated entity is permitted to access or use. \\
Identity and Access Management (IAM) & The broad framework of policies and technologies for ensuring that the right entities (users or agents) have the appropriate access to technology resources. \\
OAuth 2.1 & A modern authorization framework that allows applications to obtain limited access to user accounts. It's the foundational standard for securing agent access to APIs. \\
Model Context Protocol (MCP) & A leading protocol for connecting AI models to external tools, data sources, and resources, enabling agents to perform actions. \\
Agent-to-Agent (A2A) Protocol & A communication protocol designed to allow AI agents to engage in structured communication with other agents to complete tasks. \\
Single Sign-On (SSO) & An authentication scheme that allows a user to log in with a single set of credentials to multiple independent software systems, often used in enterprises to manage access to agent platforms. \\
SCIM (System for Cross-domain Identity Management) & A standard protocol for automating identity lifecycle management (e.g., creation, updates, de-provisioning) across different systems for both users and agents. \\
Workload Identity & A unique, verifiable identity assigned to a software application itself (like an AI agent), allowing it to authenticate to other services without using static secrets like API keys. \\
SPIFFE / SPIRE & A framework (SPIFFE) and runtime environment (SPIRE) for providing strong, verifiable, and automatically rotated workload identities to software services. \\
Delegated Identity & A durable, first-class identity for an agent that is distinct from the initiating user's identity. This allows the agent to authenticate and operate independently over long periods for asynchronous tasks. \\
Delegated Authority & An authorization model where a user explicitly grants an agent permission to act on their behalf with a specific, limited scope. \\
On-Behalf-Of (OBO) Flow & The technical pattern for implementing delegated authority, resulting in an access token that contains distinct identifiers for both the user who granted authority and the agent performing the action. \\
Impersonation & A high-risk scenario where an agent acts in a way that is indistinguishable from the user (e.g., by using the user's credentials directly), creating a gap in accountability and audit trails. \\
Client Initiated Backchannel Authentication (CIBA) & An OpenID standard that allows an agent to request user authorization asynchronously and out-of-band, ideal for long-running tasks or when a high-risk action requires explicit, non-disruptive approval. \\
Recursive Delegation & A process where an agent delegates a sub-task and a subset of its authority to another agent, creating a multi-step authorization chain. \\
Scope Attenuation & The process of progressively narrowing permissions at each step in a recursive delegation chain to enforce the principle of least privilege. \\
Revocation & The immediate invalidation of an agent's active credentials (like an access token) to terminate its current session. \\
De-provisioning & The formal and permanent removal of an agent's identity and all associated access rights from all systems, which is the final step in its lifecycle or response to a compromise. \\
Consent Fatigue & A security risk where users, overwhelmed by excessive authorization prompts from high-velocity agents, begin to reflexively approve requests reflexively, without proper review. \\
Policy Enforcement Point (PEP) & The architectural component (e.g., an API gateway) that intercepts an incoming request from an agent and enforces an authorization decision. \\
Policy Decision Point (PDP) & The centralized service that makes the authorization decision (e.g., permit or deny) based on defined policies, which is then enforced by the PEP. \\
Trust Domain & A distinct system or environment where a single authority (like one company's Identity Provider) is responsible for authenticating and authorizing users and services. Agents often need to operate across multiple trust domains. \\
Guardrails & Technical constraints or policies designed to ensure AI agents operate safely and within intended boundaries, such as by masking sensitive data or limiting resource consumption. \\
Web Bot Auth & An emerging protocol that allows a legitimate AI agent to cryptographically prove its identity within its HTTP requests, helping websites differentiate it from malicious bots. \\
\end{longtable}

\clearpage
\bibliographystyle{plainnat}
\bibliography{references}

\end{document}